# On the All-Speed Roe-type Scheme for Large Eddy Simulation of Homogeneous Decaying Turbulence

Xue-song Li

*Key Laboratory for Thermal Science and Power Engineering of Ministry of Education, Department of Thermal Engineering, Tsinghua University, Beijing 100084, PR China*

**Abstract:** As the representative of the shock-capturing scheme, the Roe scheme fails to LES because important turbulent characteristics cannot be reproduced such as the famous $k^{-5/3}$ spectral law owing to large numerical dissipation. In this paper, the Roe scheme is divided into five parts: $\xi$, $\delta U_p$, $\delta U_u$, $\delta p_p$, and $\delta p_u$, which means basic upwind dissipation, pressure-difference-driven and velocity-difference-driven modification of the interface fluxes and pressure, respectively. Then, the role of each part on LES is investigated by homogeneous decaying turbulence. The results show that the parts $\delta U_u$, $\delta p_p$, and $\delta U_p$ have little effect on LES. It is important especially for $\delta U_p$ because it is necessary for computation stability. The large numerical dissipation is due to $\xi$ and $\delta p_u$, and each of them has much larger dissipation than SGS dissipation. According to these understanding, an improved all-speed LES-Roe scheme is proposed, which can give enough good LES results for even coarse grid resolution with usually adopted reconstruction.

## 1. Introduction

Large eddy simulation (LES) becomes more and more important for unsteady turbulence computation of incompressible and compressible flows and has obtained great achievement especially for incompressible flows. Meanwhile, as one of the most important achievement of CFD, the shock-capturing scheme is developed and generally adopted for compressible flow computation. Therefore, it is necessary for LES to identify the role of intrinsic numerical dissipation of the shock-capturing scheme. Therefore, Roe scheme, as one of the most classical and popular shock-capturing schemes, had been investigated for LES of low-Mach-number homogeneous decaying turbulence (HDT) [1, 2]. The result seems disappointing because many important turbulent characteristics cannot be reproduced. For example, the famous $k^{-5/3}$ spectral sub-ranges in the self-similar decay stage by LES of the Roe scheme can only be produced in a very narrow wave number range. In fact, in high wave-number range the numerical slope of kinetic energy spectrum reaches about -5 much larger than -5/3 for all schemes in Ref. [1, 2] even for the fifth-order [1] and ninth-order [2] scheme. It means that the physical sub-grid scale (SGS) dissipation is fully immersed by the numerical dissipation, which cannot satisfy one of the following conditions for a suitable scheme for LES [1]:

(1) numerical dissipation is much lower than the physical subgrid-scale dissipation (condition (C1)),

(2) numerical dissipation is able to mimic those of a subgrid-scale (SGS) model (condition (C2))

* Corresponding author. Tel.: 0086-10-62794617; fax: 0086-10-62795946
E-mail address: xs-li@mail.tsinghua.edu.cn (X.-S. Li).

In fact, it should be noticed that the shock-capturing scheme itself cannot be directly used for incompressible flow not only LES but also general computation even for Euler flows, or else the unphysical results will be produced [3, 4]. The traditional curing method is preconditioning technology, based on which many improved schemes were developed such as preconditioned Roe [5, 6], AUSM-type [7, 8] and HLL-type [9, 10] schemes, but also suffers from some limitations. These defects are concluded by Ref. [11] as the following three problems: the non-physical behavior problem, the global cut-off problem, and the checkerboard problem.

The non-physical behaviour problem means that at low Mach number speed the solution of pressure fluctuation by the shock-capturing scheme scales with the Mach number, i.e. $p(\boldsymbol{x},t) = P_0(t) + M_* p_1(\boldsymbol{x},t)$, but the physical pressure scales with the square of the Mach number, i.e., $p(\boldsymbol{x},t) = P_0(t) + M_*^2 p_2(\boldsymbol{x},t)$, where $\boldsymbol{x}$ and $t$ means space and time, respectively. Therefore, it is obviously not correct to directly use the shock-capturing scheme for LES.

The global cut-off problem means that the local Mach number is replaced by the global reference Mach number for almost all improved shock-capturing schemes by the traditional preconditioning technology. The problem limits the ability to accurately simulate low- and high-Mach number mixed flows. For example, for a flow region shock waves coexist with incompressible flows such as the boundary layer, calculation of the incompressible region cannot benefit from the preconditioning because of the global cut-off problem and will suffer from the non-physical behaviour problem. Therefore, it is also obviously not proper to use the traditional preconditioned scheme for LES.

The checkerboard problem means the classical problem of pressure-velocity decoupling leading pressure solution with checkerboard oscillation for incompressible flows. The scheme for low-Mach number flows must have a mechanism to suppress the checkerboard problem, or else the computation will become instability and divergence. Therefore, the suppressing mechanism should also be investigated for LES because it cannot be avoided.

Above three problems can all be due to the construction of the numerical dissipation of the scheme, and then three general rules were proposed to cure the problems [11]. It provides possibility of obtaining a LES scheme that satisfies the condition (C1) or (C2) by modifying a shock-capturing scheme. Based on the premise of three general rules, however, the concrete constructing method is still needed to be discovered for satisfying LES. An effort is accordingly tried in this work by investigating the LES role of each term of the numerical dissipation of the Roe scheme and then proposing an improved Roe scheme satisfying the condition (C2).

The outline of this paper is as follows. Chapter 2 gives the governing equations and the different forms of the Roe scheme. Chapter 3 discusses the effect of each part of the Roe scheme on LES. Chapter 4 proposes an improved all-speed Roe scheme satisfying the condition (C2) for LES. Finally, Chapter 5 closes the paper with some concluding remarks.

## 2. Governing Equations and the Roe Scheme

### 2.1 Governing Equations

The governing three-dimensional Navier-Stokes (N-S) equations can be written as follows:

$$\frac{\partial \boldsymbol{Q}}{\partial t}+\frac{\partial \boldsymbol{F}}{\partial x}+\frac{\partial \boldsymbol{G}}{\partial y}+\frac{\partial \boldsymbol{H}}{\partial z}=\frac{\partial \boldsymbol{F}^v}{\partial x}+\frac{\partial \boldsymbol{G}^v}{\partial y}+\frac{\partial \boldsymbol{H}^v}{\partial z},\tag{1}$$

where $\boldsymbol{Q}=\begin{bmatrix}\rho\\\rho u\\\rho v\\\rho w\\\rho E\end{bmatrix}$ is the vector of conservation variables; $\boldsymbol{F}=\begin{bmatrix}\rho u\\\rho u^2+p\\\rho uv\\\rho uw\\u(\rho E+p)\end{bmatrix}$,

$\boldsymbol{G}=\begin{bmatrix}\rho v\\\rho uv\\\rho v^2+p\\\rho vw\\v(\rho E+p)\end{bmatrix}$, $\boldsymbol{H}=\begin{bmatrix}\rho w\\\rho uw\\\rho vw\\\rho w^2+p\\w(\rho E+p)\end{bmatrix}$ are the vectors of Euler fluxes; $\boldsymbol{F}^v$, $\boldsymbol{G}^v$, and

$\boldsymbol{H}^v$ are the vectors of viscous fluxes which are not given in detail for simplicity; $\rho$ is fluid density; $p$ is pressure; $E$ is total energy; and $u, v, w$ are the velocity components in the Cartesian coordinates $(x, y, z)$, respectively.

**2.2 Original Form of the Roe Scheme**

The classical Roe scheme can be expressed as the following general sum form of a central term and a numerical dissipation term:

$$\tilde{\boldsymbol{F}}=\tilde{\boldsymbol{F}}_c+\tilde{\boldsymbol{F}}_d,\tag{2}$$

where $\tilde{\boldsymbol{F}}_c$ is the central term and $\tilde{\boldsymbol{F}}_d$ the numerical dissipation term:

$$\tilde{\boldsymbol{F}}_{c,1/2}=\frac{1}{2}\left(\bar{\boldsymbol{F}}_L+\bar{\boldsymbol{F}}_R\right),\tag{3}$$

$$\bar{\boldsymbol{F}}=U\begin{bmatrix}\rho\\\rho u\\\rho v\\\rho w\\\rho H\end{bmatrix}+\begin{bmatrix}0\\n_x p\\n_y p\\n_z p\\0\end{bmatrix},\tag{4}$$

$$\tilde{F}_{d,1/2} = -\frac{1}{2} R_{1/2} \Lambda_{1/2} \left( R_{1/2} \right)^{-1} \left( Q_R - Q_L \right), \tag{5}$$

$$R = \begin{bmatrix} n_x & n_y & n_z & 1 & 1 \\ n_x u & n_y u - n_z & n_z u + n_y & u - n_x c & u + n_x c \\ n_x v + n_z & n_y v & n_z v - n_x & v - n_y c & v + n_y c \\ n_x w - n_y & n_y w + n_x & n_z w & w - n_z c & w + n_z c \\ n_z v - n_y w + \frac{V_M^2}{2} n_x & n_x w - n_z u + \frac{V_M^2}{2} n_y & n_y u - n_x v + \frac{V_M^2}{2} n_z & H - cU & H + cU \end{bmatrix}, \tag{6}$$

$$\Lambda^{Roe} = \begin{bmatrix} |U| & & & & \\ & |U| & & & \\ & & |U| & & \\ & & & |U-c| & \\ & & & & |U+c| \end{bmatrix}, \tag{7}$$

where $c$ is the sound speed; $V_M^2 = u^2 + v^2 + w^2$; $U = n_x u + n_y v + n_z w$ is the normal velocity on the cell interface; $H$ is total enthalpy; and $n_x$, $n_y$, and $n_z$ are the components of the face-normal vector.

## 2.3 Scalar Form of the Roe Scheme

Following Ref. [5], the numerical dissipation term of the Roe scheme in Section 2.2 can also be rewritten in the following scalar form:

$$\tilde{F}_d = -\frac{1}{2} \left\{ \xi \begin{bmatrix} \Delta \rho \\ \Delta(\rho u) \\ \Delta(\rho v) \\ \Delta(\rho w) \\ \Delta(\rho E) \end{bmatrix} + \delta p \begin{bmatrix} 0 \\ n_x \\ n_y \\ n_z \\ U \end{bmatrix} + \delta U \begin{bmatrix} \rho \\ \rho u \\ \rho v \\ \rho w \\ \rho H \end{bmatrix} \right\}. \tag{8}$$

Eq. (8) can be regarded as the uniform framework for the shock-capturing scheme [12], and the three terms on the right side have explicit physical meaning: the first term is basic upwind dissipation, the second term is a modification of the interface pressure, and the third term is a modification of the interface fluxes. Ref. [12] gives the following equations which are strictly equal to the vector form in Section 2.2.

$$\xi = |U|, \tag{9}$$

$$\delta p = -\frac{|U-c|-|U+c|}{2}c\beta + \left[|U| - \frac{|U-c|+|U+c|}{2}\right]\left[U\Delta\rho - \Delta(\rho U)\right], \quad (10)$$

$$\delta U = \frac{1}{\rho}\left(\frac{|U-c|+|U+c|}{2} - |U|\right)\beta + \frac{|U-c|-|U+c|}{2\rho c}\left[U\Delta\rho - \Delta(\rho U)\right], \quad (11)$$

where

$$\beta = \frac{\gamma-1}{c^2}\left[\frac{V_M^2}{2}\Delta\rho - u\Delta(\rho u) - v\Delta(\rho v) - w\Delta(\rho w) + \Delta(\rho E)\right] \quad (12)$$

and $\gamma$ is the ratio of the specific heat values.

With the assumption
$$\Delta(\rho\phi) = \rho\Delta\phi + \phi\Delta\rho, \quad (13)$$
where $\phi$ represents one of the fluid variables, following equations can be obtained:

$$\beta = \frac{\Delta p}{c^2} \quad \text{and} \quad U\Delta\rho - \Delta(\rho U) = -\rho\Delta U \quad (14)$$

Therefore, in fact the terms $\delta U$ and $\delta p$ can be subdivided as sub-terms driven by pressure-difference $\Delta p$ and velocity-difference $\Delta U$ as follows:

$$\delta U = \delta U_p + \delta U_u, \quad (15)$$

$$\delta p = \delta p_p + \delta p_u. \quad (16)$$

$$\delta p_p = -\frac{|U-c|-|U+c|}{2}\frac{\Delta p}{c}, \quad (17)$$

$$\delta p_u = -\left[|U| - \frac{|U-c|+|U+c|}{2}\right]\rho\Delta U, \quad (18)$$

$$\delta U_p = \left(\frac{|U-c|+|U+c|}{2} - |U|\right)\frac{\Delta p}{\rho c^2}, \quad (19)$$

$$\delta U_u = -\frac{|U-c|-|U+c|}{2c}\Delta U. \quad (20)$$

That is to say, $\delta U_p$, $\delta U_u$, $\delta p_p$, and $\delta p_u$ denote the pressure-difference-driven and velocity-difference-driven modifications on the interface velocity and pressure, respectively.

For the low-Mach number flows, Eqs. (17)-(20) can be simplified as follows:

$$\delta p_p = \frac{U}{c}\Delta p, \tag{21}$$

$$\delta p_u = (c-|U|)\rho\Delta U, \tag{22}$$

$$\delta U_p = (c-|U|)\frac{\Delta p}{\rho c^2}, \tag{23}$$

$$\delta U_u = \frac{U}{c}\Delta U. \tag{24}$$

As demonstrated in Ref. [11, 12], the non-physical behaviour problem is due to $\delta p_u$ and the checkerboard problem is due to $\delta U_p$. The Roe scheme has an inherent mechanism to suppress the checkerboard problem because $\delta U_p$ plays the role similar to the momentum interpolation method [13], but fail to the non-physical behaviour problem because the coefficient of the velocity-difference dissipation term $\Delta U$ in $\delta p_u$ is too large with the order of $\mathrm{O}(c)$. $\delta U_u$ and $\delta p_p$ seem trivial from the current viewpoint.

Above discussions mean that the numerical dissipation of the Roe scheme can be divided into five parts, i.e. $\xi$, $\delta U_p$, $\delta U_u$, $\delta p_p$, and $\delta p_u$, and indicate that these parts have different effect on computation and can be modified independently according to requirement. It provides a new way to develop LES scheme, which is different from traditional way of increasing order of scheme and seems important especially under the current condition the traditional way suffer from great difficult for LES [1, 2]. Therefore, in the following chapter the roles of five parts are investigated for LES by homogeneous decaying turbulence.

## 3. Mechanism of the Roe Scheme for LES of Homogeneous Decaying Turbulence

### 3.1 Numerical Method

In order to understand the effect of five parts of the Roe scheme on LES, nine cases are designed in Table 1. In Table 1, the number means the coefficient of each part of Roe scheme and Smagorinsky model (denoted as SMA), which is the classical SGS model of LES as follows:

$$\mu_{SMA} = \rho C_s^2 \Delta^2 \sqrt{2 S_{ij} S_{ij}}, \tag{25}$$

where $S_{ij} = \frac{1}{2}\left(\frac{\partial u_i}{\partial x_j} + \frac{\partial u_j}{\partial x_i}\right)$, the filter width $\Delta$ is chosen equal to the cell size, and the Smagorinsky constant $C_S$ is chosen equal to 0.2.

Therefore, Case 1 (denoted as Cen-SMA for clarity) is normal practice for LES, which adopts SMA and a centre scheme, i.e.

$$\tilde{\bm{F}}_d = 0. \tag{26}$$

Case 2 (denoted as Cen) adopts only the centre scheme without SGS model for understanding the behaviour of the scheme itself. Case 3 (denoted as Roe) is just the Roe scheme because all five parts are adopted. Case 4-8 (denoted as $\xi$, $\delta U_p$, $\delta U_u$, $\delta p_p$, and $\delta p_u$, respectively) adopt only one part of the Roe scheme, respectively, for investigating the role of each part. For mimicking the Smagorinsky model, Case 9 (denoted as $0.5\xi$) adopts only $\xi$ with a coefficient of 0.5, i.e.

$$\tilde{\bm{F}}_d = 0.5 \zeta \Delta \bm{Q} = 0.5 |U| \Delta \bm{Q}, \tag{27}$$

which means only half numerical dissipation of the common basic upwind dissipation in this case.

Table 1. Nine Cases

| | $\xi$ | $\delta U_p$ | $\delta U_u$ | $\delta p_p$ | $\delta p_u$ | Smagorinsky Model (SMA) |
|---|---|---|---|---|---|---|
| Case 1 (Cen-SMA) | 0 | 0 | 0 | 0 | 0 | 1 |
| Case 2 (Cen) | 0 | 0 | 0 | 0 | 0 | 0 |
| Case 3 (Roe) | 1 | 1 | 1 | 1 | 1 | 0 |
| Case 4 ($\xi$) | 1 | 0 | 0 | 0 | 0 | 0 |
| Case 5 ($\delta U_p$) | 0 | 1 | 0 | 0 | 0 | 0 |
| Case 6 ($\delta U_u$) | 0 | 0 | 1 | 0 | 0 | 0 |
| Case 7 ($\delta p_p$) | 0 | 0 | 0 | 1 | 0 | 0 |
| Case 8 ($\delta p_u$) | 0 | 0 | 0 | 0 | 1 | 0 |
| Case 9 ($0.5\xi$) | 0.5 | 0 | 0 | 0 | 0 | 0 |

For high-order accuracy of the space discretization, MUSCL (monotone upstream-centered schemes for conservation laws) [14] reconstruction with three-order interpolation, i.e. the MUSCL4 reconstruction in Ref. [1], is adopted because it is generally used and has enough accuracy compared with other methods even higher-order reconstructions [1, 2]. The limiter for shock is not used because in fact it is not necessary for low-Mach number flows. The finite volume method, rather than the finite difference method, is adopted as Ref. [2], and then the conclusions can be easily extended to practical CFD computation of engineering problems. For the time discretization, the four-stage Runge–Kutta scheme is adopted.

**3.2 LES of Homogeneous Decaying Turbulence**

The initial conditions of homogeneous decay turbulence have uniform density and temperature, and velocity field with power-law spectra $\sim k^4 e^{-2(k^2/k_0^2)}$, where $k_0 = 2$. The initial root-mean-square (RMS) Mach number is 0.2. The spatial mesh resolutions are chosen as $32^3$, $64^3$, and $128^3$, where $128^3$ grid can obtain enough good results, and $32^3$ grid is important for complex practical flows of engineering problems because limited by the computational resource the mesh resolution is usually only equivalent to $32^3$ grid and even lower in practice. All simulations are carried out up to a time of $t = 5$ for all mesh resolutions, which corresponding to approximately 7 eddy turnover times.

According to the famous Kolmogorov theory, in self-similar decay stage, energy is passed down from low wave number $k$ to high wave number with $k^{-5/3}$ spectral law in the inertial sub-range, and finally is dissipated into heat in the dissipative sub-range of sufficiently small length scales by the viscosity of the fluid. In order to test ability to produce important $k^{-5/3}$ sub-range, the kinetic energy spectrum is shown in Fig. 1 – Fig. 4. As expected, Case 1 (Cen-SMA) produces good results.

For Case 2 (Cen), Case 5 ($\delta U_p$), Case 6 ($\delta U_u$), and Case 7 ($\delta p_p$), however, they all fail to convergence for all mesh resolutions. Fig. 1 gives the kinetic energy spectrum of the four failed cases before their divergence at $32^3$ grid. It can be noticed that energy is accumulated in high wave number sub-range, which is due to lack of a dissipation mechanism whether physical or numerical. The result of Case 6 is almost overlapping with that of Case 2, which means that the numerical dissipation of the term $\delta U_u$ is almost zero. The result of Case 7 means that the term $\delta p_p$ has small negative dissipation. It is important that the term $\delta U_p$ has small dissipation based on the result of Case 5. Although for homogeneous decay turbulence the term $\delta U_p$ is not necessary, it has to be kept in the scheme to suppress the checkerboard problem for practical wall flows, or else the computation will diverge. Then, it is fortunately that the indispensable term $\delta U_p$ only has negligible dissipation for LES.

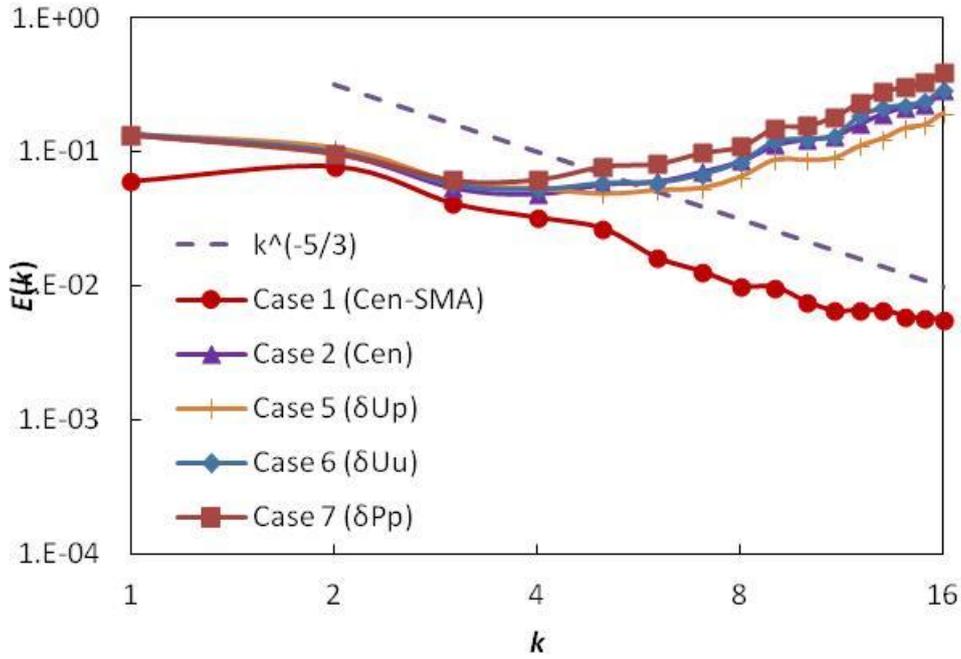

Fig.1 Kinetic energy spectrum for test Case 1, 2, 5, 6, 7 at $32^3$ grid

Fig. 2, Fig. 3 and Fig. 4 propose the results of Case 3 (Roe), Case 4 ($\xi$), Case 8 ($\delta p_u$) and Case 9 ($0.5\xi$), and similar conclusions can be obtained for different mesh resolutions. The numerical dissipation of classical Roe scheme is too large to produce correct $k^{-5/3}$ sub-range, which is replaced by approximately $k^{-5}$ sub-range for high wave number. This problem has only a little improvement by adopting higher-order reconstructions [1, 2].

The energy spectrum independently produced by the term $\xi$ and the term $\delta p_u$ also indicates that each of the terms has much larger dissipation than correct SGS model.

It should be noticed that the behaviour of the term $\delta p_u$ are not great different from that of the term $\xi$, although it does lead to the non-physical behaviour problem. The reason may be due to the one-dimensional characteristic of "homogeneity". As well known, for one-dimensional computation with the Roe scheme, the non-physical behaviour problem does not occur, and the reason is explained by Ref. [15]. For $64^3$ and $128^3$ resolution, however, the energy spectrum of $\delta p_u$ seems oscillation at high wave number. Considering the possible non-physical behaviour problem for general flows and non-monotonicity near the cut-off wave number, the term $\delta p_u$ should be reduced to zero.

Although the term $\xi$ also produces large dissipation rate, it should be kept for computation stability with some improvement reducing dissipation. Simply multiplied by 0.5, the term $\xi$ produces better energy spectrum for all resolutions as shown in following figures.

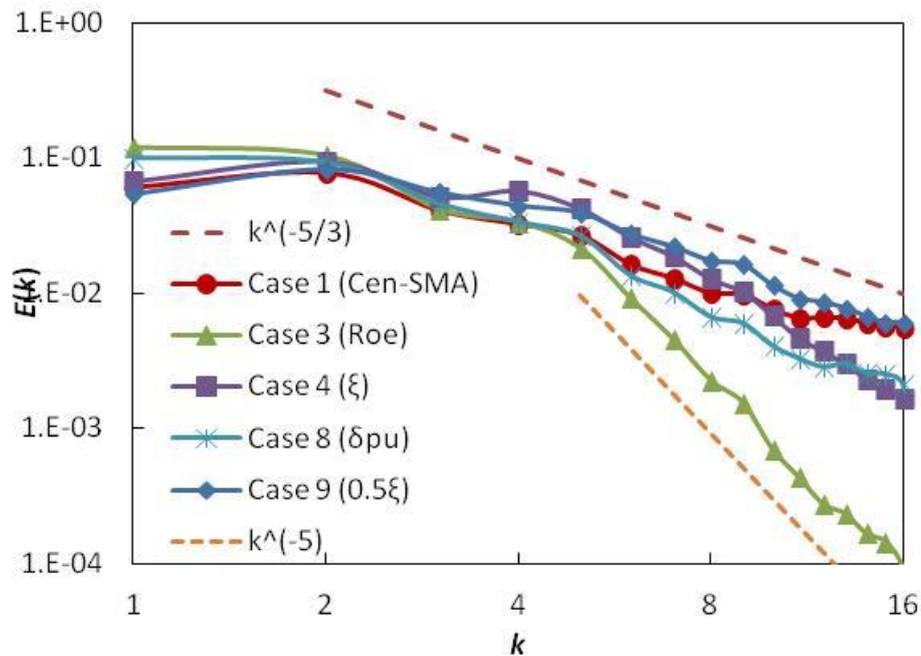

Fig.2 Kinetic energy spectrum for test Case 1, 3, 4, 8, 9 at $32^3$ grid

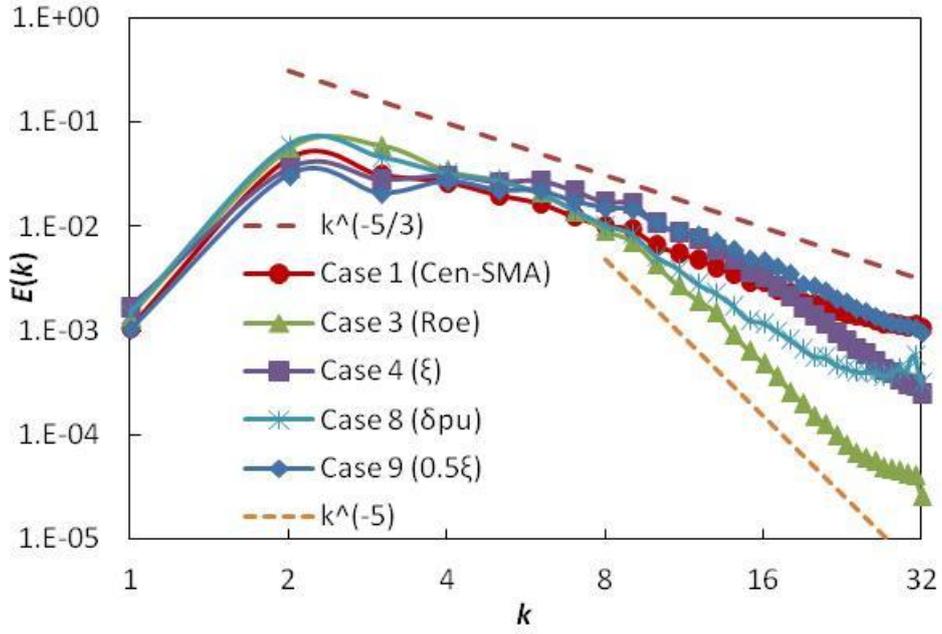

Fig.3 Kinetic energy spectrum at $64^3$ grid

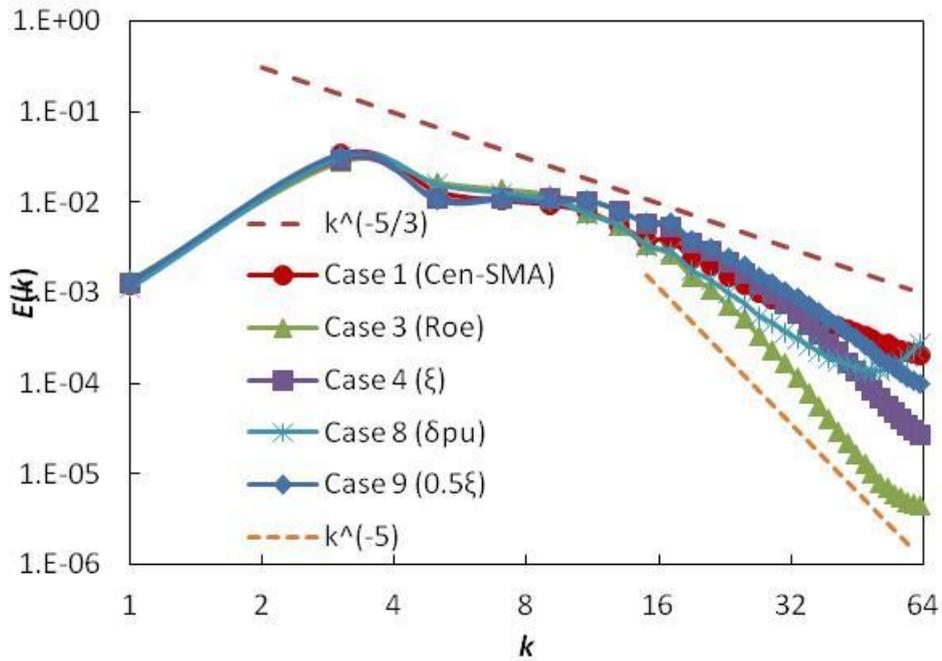

Fig.4 Kinetic energy spectrum at $128^3$ grid

Fig.5 gives iso-surfaces of vorticity which provide an intuitionistic perspective observing turbulence eddy and dissipation. Compared with Case 1 (Cen-SMA), in Case 3 (Roe) vortex tubes of small space scale corresponding to high wave number are almost disappear because of too large dissipation, and only a few large space-scale vortex tubes are produced. Obviously, Case 4 ($\xi$) is better than Case 8 ($\delta p_u$) which is better than Case 3 (Roe), and Case 9 ($0.5\xi$) seems best which is very close to Case 1 (Cen-SMA).

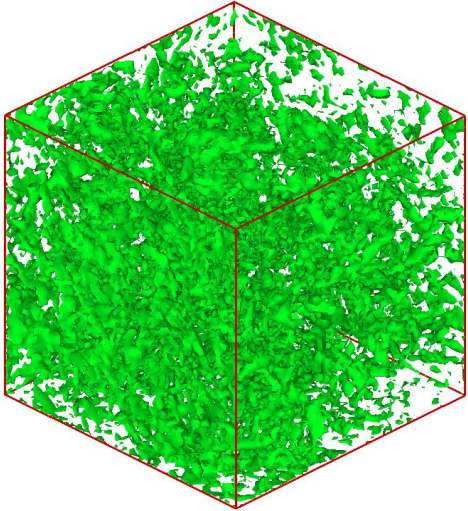
(a) Case 1 (Cen-SMA)

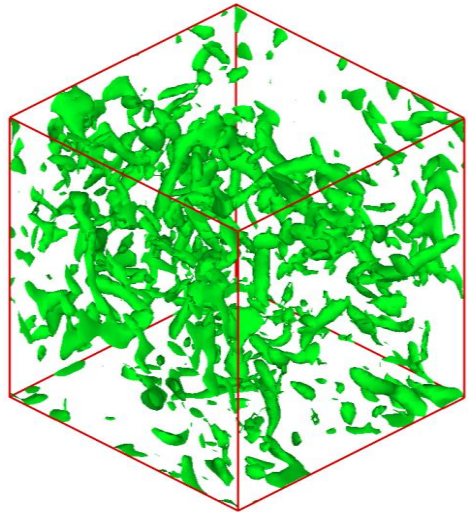
(b) Case 3 (Roe)

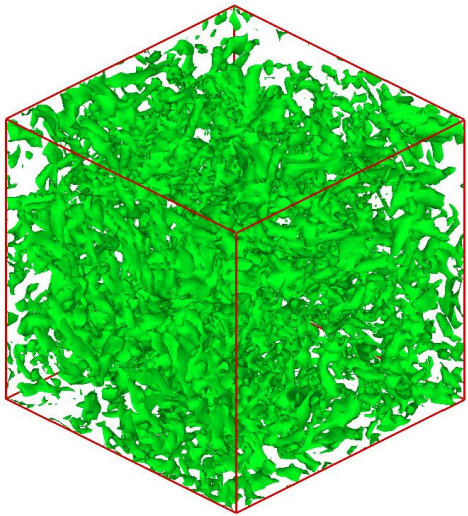
(c) Case 4 ($\xi$)

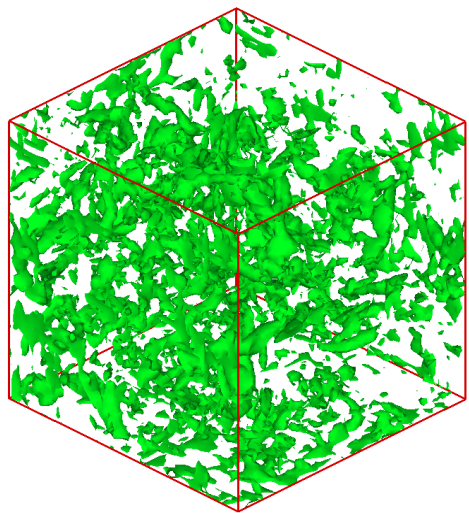
(d) Case 8 ($\delta p_u$)

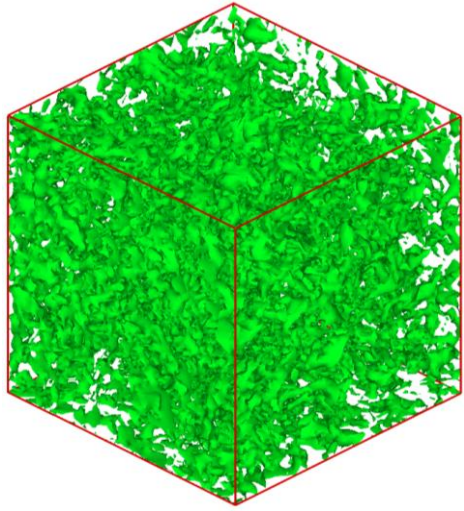

(e) Case 9 ($0.5\xi$)

Fig.5 Iso-surface of vorticity at $\omega=8.5$ in the $64^3$ grid

The resolved skewness tensor is another important parameter related to turbulence characteristic as follows:

$$Sk_{ij} = -\left\langle \left(\frac{\partial u_i}{\partial x_j}\right)^3 \right\rangle \bigg/ \left\langle \left(\frac{\partial u_i}{\partial x_j}\right)^2 \right\rangle^{\frac{3}{2}}. \tag{28}$$

Table 2. The Average of the Diagonal Components of the Resolved Skewness Tensor

|  | $32^3$ | $64^3$ | $128^3$ |
| --- | --- | --- | --- |
| Case 1 (Cen-SMA) | 0.12 | 0.24 | 0.29 |
| Case 3 (Roe) | 0.27 | 0.33 | 0.35 |
| Case 4 ($\xi$) | 0.32 | 0.34 | 0.38 |
| Case 8 ($\delta p_u$) | 0.27 | 0.31 | 0.34 |
| Case 9 ($0.5\xi$) | 0.26 | 0.28 | 0.34 |

Table 2 gives the averaged factor of the diagonal components of the resolved skewness tensor. As expected, all results increase with increasing the resolution and lie in reasonable range. However, the results also indicate that the averaged skewness factor has little relation with the numerical dissipation, because there is no obvious rule between cases.

## 4. Improvement for a LES-Roe scheme

According to the discussion in Section 3.2, an improved all-speed Roe scheme for LES is proposed as follows:

$$\xi = \alpha_1 \left[1 + f^{\alpha_2}(M)\right]|U|, \tag{29}$$

$$\delta p_u = f^{\alpha_2}(M)\left[|U| - \frac{|U-c|+|U+c|}{2}\right]\left[U\Delta\rho - \Delta(\rho U)\right], \tag{30}$$

$$\delta p_p = -\frac{|U-c|-|U+c|}{2}c\beta, \tag{31}$$

$$\delta U_u = \frac{|U-c|-|U+c|}{2\rho c}\left[U\Delta\rho - \Delta(\rho U)\right], \tag{32}$$

$$\delta U_p = \frac{1}{\rho}\left(\frac{|U-c|+|U+c|}{2} - |U|\right)\beta, \tag{33}$$

Compared with the classical Roe scheme, the differences only lie in the terms $\xi$ and $\delta p_u$ with the Mach-number-related function $f(M)$:

$$f(M) = \min\left(M\frac{\sqrt{4+(1-M^2)^2}}{1+M^2}, 1\right) \tag{34}$$

Eq. (34) is proposed [16] for smoothing transonic speed, but remains approximate $\sqrt{5}$ times dissipation in $\delta p_u$ term than that in $\xi$ term when $M \to 0$, which is not important for general low Mach number flows but has large effect on LES. Therefore, $\alpha_2$ is chosen as:

$$\alpha_2 = 4, \tag{35}$$

which make $\delta p_u$ tends to zero in the low Mach number limit as shown in Fig. 6.

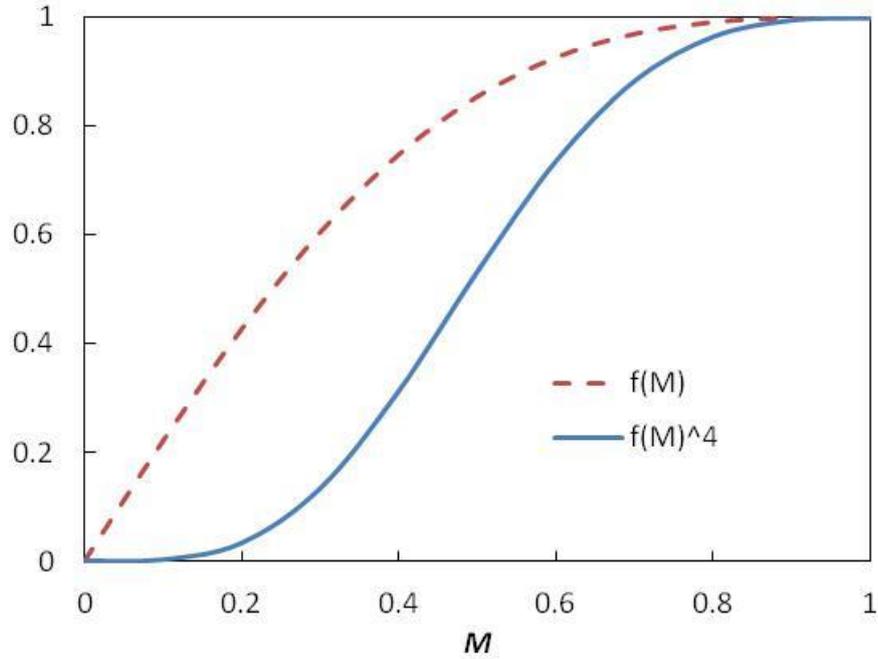

Fig.6 Effect of Mach number on the function

Based on the behaviour of Case 9 ($0.5\xi$), $\alpha_1$ is chosen as:

$$\alpha_1 = 0.5. \tag{36}$$

Therefore, Eq. (29) tends to Case 9 ($0.5\xi$) in the low Mach number limit.

The improved LES-Roe scheme Eqs. (29)-(36) satisfies the rules [11] for the all-speed scheme to overcome the non-physical behavior problem, the global cut-off problem, and the checkerboard problem. Further, as expected, the results produced by the improved LES-Roe scheme are very similar to that by Case 9 ($0.5\xi$). Fig. 7 shows that the LES-Roe scheme can produce good energy spectrum in general for all grid resolution. It means that for coarse resolution this scheme can also produce enough good results for LES computation of engineering problems.

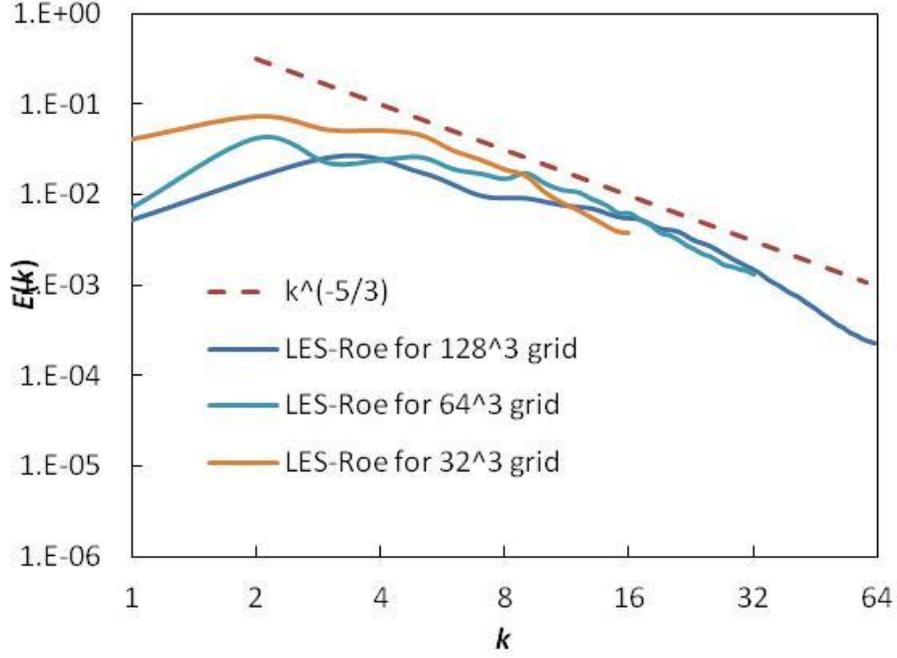

Fig.7 Kinetic energy spectrum of LES-Roe scheme at different resolution

## 5. Conclusions

In this paper, the Roe scheme is divided into five parts and then their influences on LES are investigated through homogeneous decaying turbulence as follows:

(1) The terms $\delta U_u$, $\delta p_p$, and $\delta U_p$ has little numerical dissipation to affect LES. This conclusion is important especially for the term $\delta U_p$ because it is necessary to suppress the checkerboard problem for computation of general flows. Now the term $\delta p_u$ can be kept in the scheme for computational stability without worrying about its influence of dissipation.

(2) Each of terms $\xi$ and $\delta p_u$ has much larger numerical dissipation than SGS dissipation. Because the term $\delta p_u$ can be set to zero in low Mach number limit, only the term $\xi$ needs to be focused. The improvement for $\xi$ should mimic SGS model at the same time to maintain stability. From current results, simply multiplying $\xi$ by a coefficient of 0.5 seems achieve this goal.

Based on above understanding, a LES-Roe scheme is proposed. This scheme means that it is possible to carry out LES with relatively coarse grid resolution and usually adopted schemes improved by some little modification.

**Acknowledgments**

This work is supported by Project 51276092 of the National Natural Science Foundation of China. The author also thanks Prof. Li XinLiang for helping of spectral analysis.

**References**

[1] E. Garnier, M. Mossi, P. Sagaut, P. Comte, and M. Deville. On the Use of Shock-Capturing Schemes for Large-Eddy Simulation, Journal of Computational Physics 153 (1999) 273-311.

[2] B. Thornber, A. Mosedale, D. Drikakis. On the Implicit Large Eddy Simulation of homogeneous decaying turbulence, Journal of Computational Physics 226 (2007) 1902-1929.

[3] H. Guillard, C. Viozat, On the Behaviour of Upwind Schemes in the Low Mach Number Limit, Computers and Fluids 28 (1999) 63-86.

[4] E. Turkel, Preconditioning Techniques in Computational Fluid Dynamics, Annual Reviews of Fluid Mechanics 31 (1999) 385-416.

[5] J.M. Weiss, W.A. Smith, Preconditioning Applied to Variable and Const Density Flows, AIAA Journal 33 (1995) 2050-2057.

[6] E. Turkel, Preconditioning Techniques in Computational Fluid Dynamics, Annual Reviews of

Fluid Mechanics 31 (1999) 385-416.

[7] M.S. Liou, A Sequel to AUSM, Part II: AUSM+-up for All Speeds, Journal Computational Physics 214 (2006) 137–170.

[8] E. Shima, K. Kitamura, Parameter-Free Simple Low-Dissipation AUSM-family Scheme for All Speeds AIAA Journal 49 (2011) 1693–1709.

[9] S.H. Park, J.E. Lee, J.H. Kwon. Preconditioned HLLE Method for Flows at All Mach Numbers. AIAA Journal 44 (2006) 2645–2653.

[10] H. Luo, J.D. Baum, Extension of Harten-Lax-van Leer Scheme for Flows at All Speeds, AIAA Journal 43 (2005) 1160–1166.

[11] X.S. Li, C.W. Gu, Mechanism of Roe-type Schemes for All-Speed Flows and Its Application, Computers and Fluids 86 (2013) 56–70.

[12] X.S. Li, Uniform Algorithm for All-Speed Shock-Capturing Schemes, International Journal of Computational Fluid Dynamics 28 (2014) 329–338.

[13] X.S. Li, C.W. Gu, The Momentum Interpolation Method Based on the Time-Marching Algorithm for All-Speed Flows, Journal of Computational Physics 229 (2010) 7806–7818.

[14] B. Van Leer, Towards the Ultimate Conservative Difference Scheme. V. A Second-Order Sequel to Godunov's Method, Journal of Computational Physics 32 (1979) 101-136.

[15] H. Guillard, On the behavior of upwind schemes in the low Mach number limit. IV: P0 approximation on triangular and tetrahedral cells, Computers & Fluids 38 (2009) 1969–1972.

[16] X.S. Li, C.W. Gu, J.Z. Xu, Development of Roe-Type Scheme for All-Speed Flows Based on Preconditioning Method, Computers and Fluids 38 (2009) 810-817.